\documentclass[aps,prb, 
superscriptaddress,
reprint,
longbibliography,
floatfix]{revtex4-1}
\usepackage{graphicx}
\usepackage{dcolumn}
\usepackage{bm}
\usepackage{textcomp}
\usepackage{amsmath}
\usepackage{amssymb}
\usepackage{epstopdf}
\usepackage{array,multirow}
\usepackage{mathtools}

\newcommand\Tstrut{\rule{0pt}{2.6ex}}

\newcommand{\rv}{{\bf r}}
\newcommand{\kv}{{\bf k}}
\newcommand{\qv}{{\bf q}}
\newcommand{\vv}{{\bf v}}
\newcommand{\ep}{{\epsilon}}
\newcommand{\ql}{{q_\lambda}}
\newcommand{\qlq}{{q_\lambda(q)}}
\newcommand{\qpql}{{q+q_\lambda}}
\newcommand{\rAAi}{{\rm \AA}^{-1}}
\newcommand{\srAAi}{\:{\rm \AA}^{-1}}
\newcommand{\bra}[1]{{\langle #1 |}}
\newcommand{\ket}[1]{{| #1 \rangle}}
\newcommand{\braket}[2] {{\langle #1 | #2 \rangle}}

\DeclarePairedDelimiter\abs{\lvert}{\rvert}%

\makeatletter
\let\oldabs\abs
\def\abs{\@ifstar{\oldabs}{\oldabs*}}


\preprint{}
\begin{document}

\title[ ]{Charged impurity scattering in two-dimensional materials with ring-shaped valence bands: GaS, GaSe, InS, and InSe}

\author{Protik Das}
\affiliation{%
Department of Electrical and Computer Engineering, University of California, Riverside, California 92521, USA
}%

\author{Darshana Wickramaratne}%
\affiliation{%
Materials Department, University of California, Santa Barbara, California 93106, USA}

\author{Bishwajit Debnath}
\affiliation{%
Department of Electrical and Computer Engineering, University of California, Riverside, California 92521, USA
}%

\author{Gen Yin}
\affiliation{%
Department of Electrical and Computer Engineering, University of California, Riverside, California 92521, USA
}%

\author{Roger K. Lake}
\affiliation{%
Department of Electrical and Computer Engineering, University of California, Riverside, California 92521, USA
}%

\date{\today}

\begin{abstract}
The singular density of states and the two Fermi wavevectors resulting from a 
ring-shaped or ``Mexican hat'' valence band give rise to unique trends
in the charged impurity scattering rates and charged impurity limited mobilities.
Ring shaped valence bands are common features of many monolayer and few-layer
two-dimensional materials including the III-VI materials GaS, GaSe, InS, and InSe. 
The wavevector dependence of the screening, calculated within the random phase
approximation, is so strong that it is the dominant factor determining
the overall trends of the scattering rates and mobilities 
with respect to temperature and hole density.
Charged impurities placed on the substrate and in the 2D channel are considered.
The different wavevector dependencies of the bare Coulomb potentials
alter the temperature dependence of the mobilities.
Moving the charged impurities 5 {\AA} from the center of the channel to the substrate
increases the mobility by an order of magnitude.
\end{abstract}

\pacs{Valid PACS appear here}
\maketitle

\section{\label{sec:introduction}Introduction}

Atomically thin two-dimensional (2D) materials are being investigated 
for a range of applications including emerging beyond-CMOS electronic devices,
thermoelectrics, and optoelectronics.  
%
%
%
A number of these materials have ``ring-shaped'' valence bands.
These materials include the
semiconducting III-VI monochalcogenides, GaS, GaSe, InS, and InSe
\cite{darshanamhat, zolyomi_GaX,zolyomi_InX, GaS_photodetector_AnPingHu, Hennig_GroupIII_ChemMat, SGLouie_GaSe_arxiv,	WYao_GaS_GaSe_arxiv, guo2017band}, 
bilayer graphene when subject to a vertical electric field \cite{Fermi_ring_Neto_PRB07,Falko_BLG_Lifshitz_PRL14, MacDonald_bi_gap_PRB07},
monolayers of Bi$_2$Se$_3$\cite{darshanamhat} and Bi$_2$Te$_3$\cite{Zahid_Lake,Lundstrom_Jesse_Bi2Te3,Udo_Bi2Se3},
few-layers of Bi$_2$Se$_3$ intercalated with 3d transition metals\cite{li2016gate},
monolayer SnO\cite{seixas2016multiferroic, houssa2017hole},
2D hexagonal lattices of group-VA elements \cite{sevinccli2017quartic},
and hexagonal group-V binary compounds\cite{nie2017room}.

A ring-shaped valence band edge results in a $1/\sqrt{E}$ singularity in the 2D density of states
and a step function turn on of the density of modes at the valence band edge
\cite{Lundstrom_Jesse_Bi2Te3,darshanamhat,GaSe_Ajayan_NL13,Ajayan_InSe,GaSe_Geohagen_ACSNano}.
At low temperatures, density functional theory calculations show that the singularity in the
density of states leads to a ferromagnetic phase transition at sufficient hole 
doping in GaS and GaSe\cite{SGLouie_GaSe_arxiv,	WYao_GaS_GaSe_arxiv}.
More recent calculations find that such a transition is a general property of the
Mexican hat dispersion \cite{seixas2016multiferroic}.

The ring-shaped dispersion affects ionized impurity scattering through the density of states,
the momentum transfer required to scatter around the ring, and the
momentum dependence of the screening.
The question we address is what is the influence of the ``ring-shaped'' dispersion
on the temperature, density, and Fermi energy dependence of the ionized 
impurity scattering rates and ionized impurity limited mobility.

Prior studies have theoretically investigated the role of ionized impurity scattering
in two--dimensional materials with a parabolic dispersion.  
Ionized impurity scattering can severely limit the mobility 
in the transition metal dichalcogenides such as
MoS$_{2}$ \cite{DJena_PRX} and 
give rise to an unexpected temperature 
dependence of the mobility \cite{ong2013mobility}. 
It has been predicted that reducing the doping can enhance the linear screening
response within the Thomas-Fermi theory \cite{kolomeisky2016anomalous}.
%
%
%
%
The role of screening on charged
impurity scattering and charged impurity limited mobility 
in materials with a ring-shaped dispersion
has not yet been addressed.  

We address this question
using an analytical bandstructure model with parameters extracted from first principles calculation.  
Screening is included within the random phase approximation.
Polarization functions and scattering rates are analyzed,
and the ionized impurity limited hole mobility of
the III-VI materials, GaS, GaSe, InS, and InSe, are compared.

\section{\label{sec:theory}Theory}
\begin{figure}
	\includegraphics[width=\linewidth,keepaspectratio]{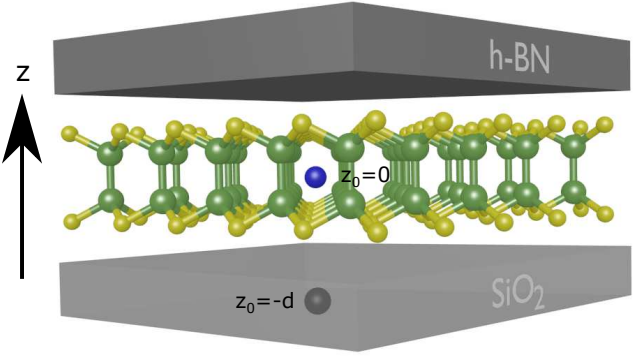}     
	\caption{\label{fig:device} 
	Monolayer GaS between SiO$_2$ substrate and BN capping layer.
	The black atom depicts an ionized impurity inside the capping layer
	5.5{\AA} from the channel.
	And the blue atom depicts a ionized impurity inside the monolayer
	GaS.	
	}
\end{figure}

The materials and geometry of the problem consist of a monolayer 2D semiconducting material 
on a insulating substrate encapsulated by an insulating capping layer
which could be the same as the substrate. 
Example insulating materials are BN or SiO$_2$.
The structure is illustrated in Fig. \ref{fig:device} with SiO$_2$ for the 
substrate and BN for the capping layer.
A cylindrical coordinate system is used with $\rv$ a vector in the $x$--$y$ plane.
The origin is at the center of the semiconductor.
Charged impurities will be considered for two different positions, in the center of the 2D semiconductor, $z=0$, and
on the surface of the substrate, $z=-d$.
Accounting for the $5$ {\AA} thickness of a monolayer III-VI semiconductor and the $3$ {\AA} van der Waals gap,\cite{darshanamhat} 
we use $d=5.5$ {\AA} for the charged impurities on the surface of the substrate.
The value of the impurity density used in all calculations is $10^{12}$ cm$^{-2}$. 
All calculated scattering rates are linearly proportional to the impurity density,
and all mobilities are inversely proportional to the impurity density, so any
calculated values can be scaled for different impurity densities.

\begin{figure}
	\includegraphics[width=\linewidth,keepaspectratio]{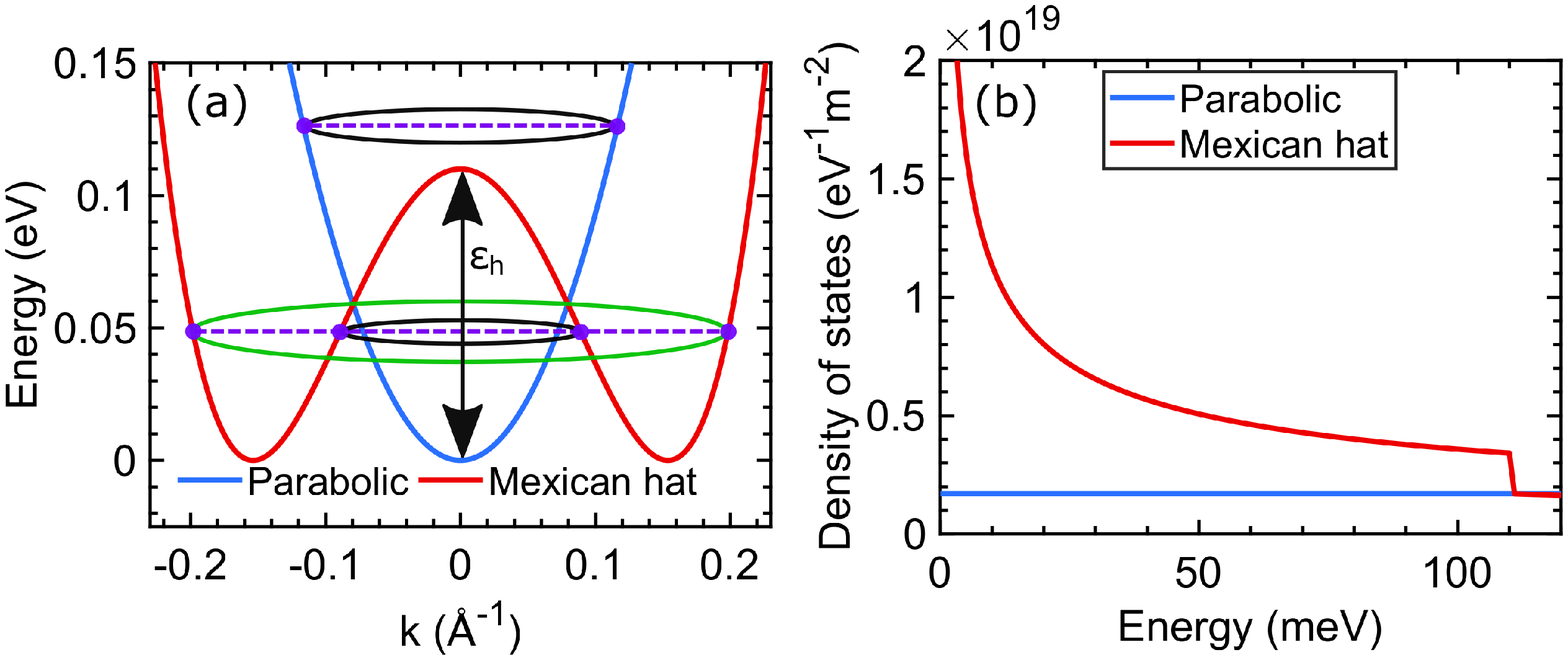}     
	\caption{\label{fig:dispersion_dos} 
		(a) Comparison of a parabolic (blue) and Mexican hat dispersion (blue).
		The height of the Mexican hat band at $k=0$ is $\epsilon_{h} = 0.11$ eV.
		(b)  Density of states of the parabolic band (blue) and Mexican hat 
		dispersion (red). 
		The parabolic and Mexican hat dispersion both have an effective mass of 
		$0.409$ m$_0$.
		}                
\end{figure}

The investigation of the effect of the Mexican hat dispersion on screening, scattering,
and mobility, begins with the model quartic dispersion 
\begin{equation}
E(k) = \epsilon_h -\frac{\hbar^{2}k^{2}}{2m^*} + \frac{1}{4\epsilon_{h}}\left(\frac{\hbar^{2}k^{2}}{2m^*}\right)^2 .
\label{eq:MH}
\end{equation}
Quartic models have been previously used to investigate interactions in 
biased bilayer graphene \cite{Fermi_ring_Neto_PRB07}, multiferroic 2D materials \cite{seixas2016multiferroic},
and electronic and thermoelectric properties of group III-VI and group VA 2D 
materials \cite{darshanamhat,sevinccli2017quartic}. 
We define our momentum-energy relation such that
the hole kinetic energy is positive,
the valence band edge is at $E=0$, and
negative energies correspond to
energies in the band gap. 
The term $\ep_h$ in Eq. (\ref{eq:MH}) is the height of the `hat' at $k=0$ and $m^*$ is the magnitude of the
effective mass at $k=0$ (the top of the hat).
The addition of the constant term $\ep_h$ in Eq. (\ref{eq:MH}), shifts the dispersion
so that the minimum energy, corresponding to the band edge, occurs at $E=0$.
For energies $0 < E < \ep_h$ the Mexican hat dispersion has two Fermi 
wavevectors
corresponding to the two branches of the dispersion.
In this energy region, the Fermi surface consists of two concentric circles 
shown in Fig. \ref{fig:dispersion_dos}(a).
The radii of the two circles are
$k_1 = \frac{\sqrt{2 m^* \ep_h}}{\hbar} \sqrt{1 - \sqrt{E/\ep_h} }$
and 
$k_2 = \frac{\sqrt{2 m^* \ep_h}}{\hbar} \sqrt{1 + \sqrt{E/\ep_h} }$.
At the band edge, $E=0$, the two circles merge into a single circle with a radius
of $k_0 = 2 \sqrt{m^*\ep_h} / \hbar$.
The effective mass at the band edge determined from 
$\frac{1}{m^*(k_0)} = \left. \frac{\partial^2 E}{\hbar^2 \partial k^2} \right|_{k=k_0}$
is $m^*/2$.
The single--spin densities of states for each individual $k$-space ring are identical
and equal to 
\begin{equation}
\begin{array}{ll}
D_{1}(E) = D_2(E) = \frac{m^*}{2 \pi \hbar^2} \sqrt{ \frac{\ep_h}{E} } \;\;\;\;\;\;\;
&
(0 \leq E \leq \ep_h),
\end{array}
\end{equation}
The total single-spin density of states is given by the sum and is equal to
\begin{equation}
D(E)=
\left\{
\begin{array}{ll}
\frac{m^*}{\pi \hbar^2} \sqrt{ \frac{\ep_h}{E} } \;\;\;\;
&
(0 \leq E \leq \ep_h)\\ 
\frac{m^*}{2 \pi \hbar^2 } \sqrt{ \frac{\ep_h}{E} }
&
\left( \ep_h < E \right) \; .
\end{array}
\right. 
\label{eq.DE_hat}
\end{equation}
The density of states is plotted in Fig. \ref{fig:dispersion_dos}(b) using $m^* = 0.409$ m$_0$ and $\ep_h = 0.11$ eV,
which are similar to the values for monolayer GaS \cite{darshanamhat}.
The density of states diverges as $1/\sqrt{E}$ at the band edge, and it is equal to the 
single--spin parabolic density of states, $\frac{m^*}{2\pi \hbar^2}$, at the top of the hat.

A parabolic dispersion $E(k) = \frac{\hbar^{2}k^{2}}{2m^*}$ will be used as a reference and for comparison.
The parabolic and Mexican hat dispersions and density of states are plotted together in Fig. \ref{fig:dispersion_dos}.
An effective mass of $m^* = 0.409 m_0$ is used for both dispersions.

\begin{figure}
	\includegraphics[width=\linewidth,keepaspectratio]{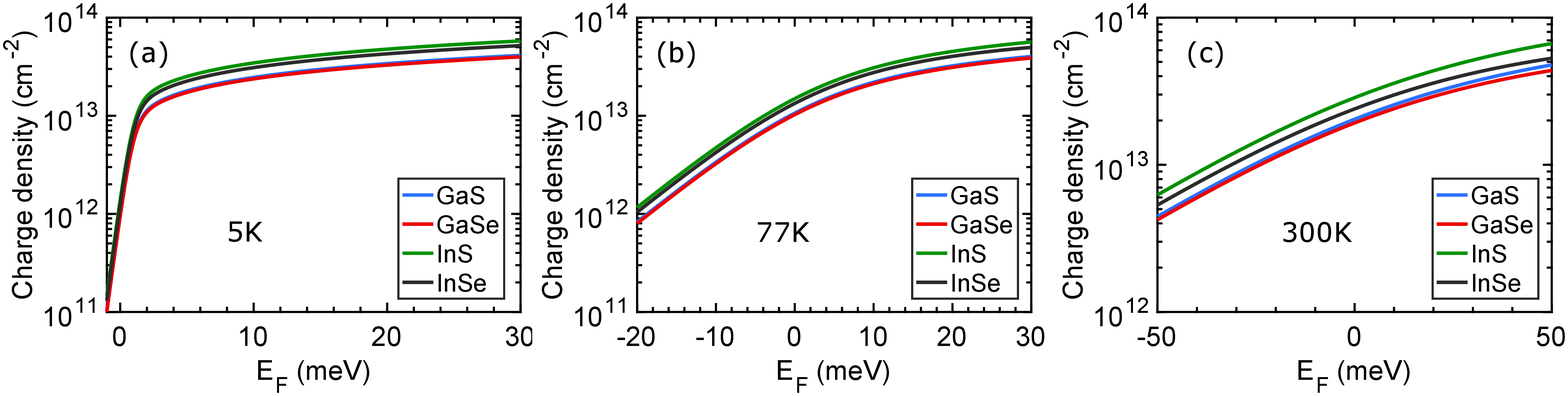}
	\caption{\label{fig:charge_density} 
		Carrier concentration of GaS, GaSe, InS and InSe as a function of Fermi 
		level $E_F$ for (a) $5$ K, (b) $77$ K and (c) $300$ K.
		Parameters used for materials are tabulated in	Table 
		\ref{tab:mat_params}.
	}                
\end{figure}

The two--dimensional Fourier transform of the bare Coulomb potential for a point charge at position 
$x = y = 0$, $z = z_0$ is
\begin{equation}
v(q) = \frac{e^2 e^{-q|z-z_0|}}{2\epsilon q},
\label{eq:vqbare}
\end{equation}
where $e$ is the charge of electron, $\epsilon$  is the average static dielectric constant
and $q$ is the momentum transfer.
Since all of the III-VI materials have relative dielectric constants in the range of
$3 - 4$, we will use the dielectric constant of the semiconductor.
Within the random phase approximation,
the screened Coulomb potential is
\begin{equation}
V(q,z) = \frac{v(q)}{1 - \Pi(q) v(q)} .
\label{eq:coulomb_pot}
\end{equation}
Substituting Eq. (\ref{eq:vqbare}) into Eq. (\ref{eq:coulomb_pot}) gives
the 2D RPA screened potential,
\begin{align}
V(q) &= \frac{e^{2}}{2\epsilon (q e^{q|z-z_0|} - \frac{e^2}{2\epsilon}\Pi (q))}
\nonumber \\
&= \frac{e^{2}}{2\epsilon (q e^{q|z-z_0|} + \qlq)},
\label{eq:coulomb_pot2}
\end{align}
where $\qlq \equiv -\frac{e^2}{2\epsilon}\Pi (q)$ is the wavevector dependent
inverse screening length.
In the static limit, the polarization function is 
\cite{maldague1978many}
\begin{equation}
\Pi (q) = \frac{2}{A} \sum_{\bf{k}} 
\frac{ f(E_{\kv + \qv}) - f(E_{\kv}) }
{E_{\kv + \qv} - E_{\kv}}
\label{eq:Pi}
\end{equation}
where $A$ is the area, $E_\kv$ is the eigenenergy at wavevector $\kv$, and
$f(E)$ is the Fermi-Dirac function.
The factor of 2 is for spin degeneracy, since the Mexican hat bands in the III-VI materials
are spin degenerate.
For both the Mexican hat and parabolic dispersions, $E_\kv$ is only a function of the magnitude of $k$.
Therefore, we define the variable,
\begin{equation}
	k_+ = \left|\bf{k}+\bf{q}\right| = \sqrt{k^2 + q^2 + 2kq\cos{\theta}},
\end{equation}
and calculate the polarization from Eq. \ref{eq:Pi},
\begin{equation}
	\Pi (q) = \frac{1}{2\pi^2} \int_{0}^{\infty} dkk
	\int^{2\pi}_0 d\theta \frac{f(E(k_+)) - f(E(k))}
{E(k_+) - E(k)} .
\end{equation}
In the limit $\qv \rightarrow 0$,
the polarization function becomes the negative of the thermally averaged density
of states at the Fermi level,
\begin{equation}
\Pi (q=0) = \int_{0}^{\infty} dE D(E) \frac{\partial f}{\partial E} ,
\label{eq:pi_analytical}
\end{equation}
where $D(E)$ is the density of states.
Using the $\qv \rightarrow 0$ limit for $\Pi(\qv)$ in Eq. (\ref{eq:coulomb_pot2}),
gives the Thomas-Fermi form of the 2D screened Coulomb potential with an inverse
screening length of $\frac{e^2}{2\ep} D(E_F)$.
For the Mexican hat dispersion
this is problematic, since
the density of states diverges near the band edge.
Note that in defining the polarization function in Eq. (\ref{eq:Pi}), $\Pi < 0$. 
%
%

To calculate the momentum relaxation time, we need the matrix elements of the 
RPA Coulomb potential.
We assume separable wavefunctions of the form 
%
$\braket{\rv}{\kv} = \frac{1}{\sqrt{A}} e^{i\kv \cdot \rv} \sqrt{\delta(z)}$
%
and take the matrix elements of $\tilde{V}(\rv) = \int \frac{d^2q}{4\pi^2} V(q) e^{i\qv \cdot \rv}$
to obtain
%
$\bra{\kv} \tilde{V} \ket{\kv'} \equiv V_{\kv,\kv'} = \frac{1}{A} V(|\kv - \kv'|)$.
%
The Fermi's golden rule expression for the inverse momentum relaxation time is given by 
\begin{equation}
\frac{1}{\tau (k)} = 
\frac{N_I 2\pi}{\hbar} 
\sum_{\kv'} 
\lvert V_{\kv',\kv} \rvert^2 \delta(E_{\kv'}-E_{\kv})
\left( 1 - \frac{\vv(\kv) \cdot \vv(\kv')}{\left| \vv(\kv) \right| ^2} \right),
\label{eq:Fermi_Au}
\end{equation}
%
%
where $N_I$ is the number of charged impurities.
For the Mexican hat dispersion, the group velocity $\vv$ is opposite to the direction
of $\kv$ on the inner ring and parallel to $\kv$ on the outer ring.
On a given branch of the Mexican hat dispersion, $E(\kv)$ is only a function of the magnitude
of $\kv$.
Therefore, by converting the sum over $\kv'$ into an integral and explicitly keeping track of the 
two branches of the dispersion, Eq. (\ref{eq:Fermi_Au}) becomes
\begin{equation}
\frac{1}{\tau(k)} = 
\frac{n_I e^4}{4 \ep^2 \hbar}
\sum_{\nu=1}^2 
D_\nu (E)
\int_0^{2\pi} \! \! \! \! d\theta \: 
\frac{ \left( 1 - \frac{\bf{v(k'_\nu)}\cdot v(\bf{k})}{v^2(k)} \right) }
{\left(q \: e^{qz_0} + \qlq \right)^2} 
\label{eq:tau}
\end{equation}
where the sum is over the two Fermi rings,
$q=\left| \kv'_\nu -\kv \right| = \sqrt{{k'_\nu }^{2}+ k^2 - 2k'_\nu k \cos\theta}$,
$k$ and $k'_\nu$ correspond to the radii of the concentric iso-energy rings in Fig. \ref{fig:dispersion_dos},
$D_\nu (E)$ is the final single--spin density of states corresponding to 
ring $\nu$,
$\vv(\kv'_\nu)$ is the final group velocity of
ring $\nu$,
and $n_I$ is the impurity density per unit area.
The value of $z_0$ is either zero for impurities placed at the center of the semiconducting monolayer
or $5.5$ {\AA} for impurities placed on the substrate.

The last term on the right of Eq. (\ref{eq:Fermi_Au}) is $1 - \frac{v'}{v}\cos(\theta_{\vv,\vv'})$
where $\theta_{\vv,\vv'}$ is the angle between 
the group velocity of state $\kv$ and the group velocity of state $\kv'$.
This term is the relative change in the component of the velocity that is parallel to the initial 
velocity $v$.
When the final velocity $v'$ is in the same direction and greater than the initial velocity $v$, 
then scattering from $v$ to $v'$ causes the carrier to speed up and 
gives a negative contribution to the momentum relaxation time \cite{1986_Neg_tau_m}.
This situation occurs for carriers that are initially near the top of the hat in Fig. \ref{fig:dispersion_dos}(a)
and then scatter to the outer ring.
However, the negative values are restricted to a range of angles centered around $180^\circ$,
and the integral over $\theta$ in Eq. (\ref{eq:tau}) is always positive.

The carrier mobility is determined from the 
average group velocity driven by an external
electric field oriented in the $x$--direction.
To linear order, this is
\begin{equation}
\langle v_x \rangle = \frac{\sum_{\bf k} v_x (\kv ) f_A ( \kv )} {\sum_{\kv} f_0 (\kv)},
\label{eq:vx}
\end{equation}
where $ f_A (\bf k) $ is the asymmetric component of the non-equilibrium distribution function.
Within the relaxation time approximation, the asymmetric distribution function can be written as,
\begin{equation}
f_A(\kv) = - \tau (k) \frac{e {\cal E}_x}{\hbar}\frac{\partial f_0(\kv)}{\partial k} \cos \theta,
\label{eq:fA}
\end{equation}
where $ f_0 (\kv) $ is the equilibrium Fermi function, ${\cal E}_x $ is the electric field along
the transport direction and $ \theta $ is the direction of $\kv$ with respect to the $k_x$ axis.
The mobility is directly evaluated from its definition, 
$\mu = \langle v_x \rangle / {\cal E}_x$.
Substituting (\ref{eq:fA}) into (\ref{eq:vx}), the final expression for carrier mobility is
\begin{equation}
\mu = - \frac{e}{2 \pi \hbar^2 p}
\int_{0}^{\infty} \! \! \! dk
\: k \: \tau(k) \, \frac{\partial f_0}{\partial \epsilon} \left(\frac{\partial \epsilon}{\partial k}\right)^2,
\label{eq:mu}
\end{equation}
where the spin--degenerate 2D hole density $p = \frac{2}{A}\sum_\kv f_0(\kv)$.

\section{\label{sec:results}Results}
The polarization function $\Pi(q)$ gives 
wavevector dependent screening.
In a two dimensional material with parabolic dispersion, 
the density of states is constant which results 
in a constant polarization function for $q<2k_F$ at low temperature.
In a Mexican hat dispersion, 
the singular density of states gives 
a strong wavevector dependence
to the polarization function at low temperature. 
It also increases the overall magnitude of the polarization function. 
The wavevector dependent inverse screening length $\qlq$
is added to the momentum transfer $q$
in the denominator of Eq. (\ref{eq:coulomb_pot2}),
and the sum determines the magnitude and wavevector dependence
of the screened Coulomb interaction.
Therefore, we begin by analyzing $q_\lambda$ as a function of $q$ 
for the Mexican hat dispersion.

\begin{figure}
\includegraphics[width=\linewidth,keepaspectratio]{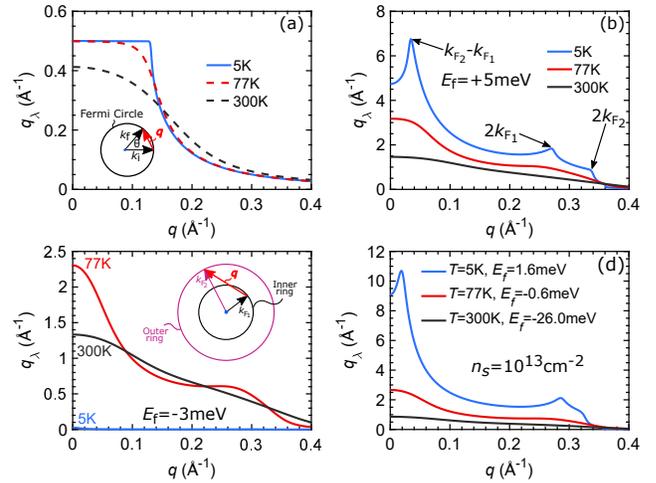}
\caption{
\label{fig:pi} 
(a)--(c) $q_\lambda(q) \equiv -\frac{e^2}{2\ep} \Pi(q)$ 
at three different temperatures: 5 K, 77 K and 300 K.
(a) $\qlq$ for a parabolic dispersion with $E_F = 40$ meV.
(b) $\qlq$ for a Mexican hat dispersion with $E_F = +5$ meV. 
(c) $\qlq$ for a Mexican hat dispersion with $E_F = -3$ meV.
The inset shows the two iso-energy rings in momentum space of the Mexican hat dispersion. 
The momentum transfer $q$ between two rings is shown.
(d) $\ql(q)$ for a Mexican hat dispersion for a fixed charged density of $10^{13}$ cm$^{-2}$.
The corresponding Fermi energies at each temperature are shown on the graph.
For both the parabolic and Mexican hat dispersions, the band structure
parameters are the same as those used in Fig. \ref{fig:dispersion_dos},
and the dielectric constant is $\ep = 3.1 \ep_0$. 
}                
\end{figure}

To provide a point of reference,
we first show in Fig.~\ref{fig:pi}(a) the well--known 
wavevector dependent inverse screening length $q_\lambda$
resulting from a parabolic dispersion
with the Fermi level fixed at $40$ meV above the band edge.
At low temperature and for wavevectors smaller than $2 k_F$, 
the magnitude of $q_\lambda$ is simply 
$\frac{e^2}{2 \epsilon} \: \frac{m^*}{\pi \hbar^2}$,
i.e. $\frac{e^2}{2 \epsilon}$ times the density of states at the Fermi level.
This is equal to $3.78$  $m_r / \ep_r = 0.499 \; {\rm \AA}^{-1}$ where $m_r = 0.409$ is the relative effective mass
and $\ep_r = 3.1$ is the relative dielectric constant.
Since the density of states is constant, the resulting inverse screening length is 
constant up until the momentum transfer is greater than $2 k_F$.
At higher temperatures, the polarization function can be written as a convolution of the
zero--temperature polarization and a thermal broadening function \cite{maldague1978many}.
The result is that the sharp $q$-dependent features become smeared out at finite temperatures. 

Unlike the parabolic dispersion 
where scattering occurs within a single Fermi ring, 
Coulomb scattering in a Mexican hat dispersion occurs within 
and between two concentric rings for energies up to $\epsilon_h$, which
defines the height of the Mexican hat dispersion. 
Furthermore, the density of states is singular
at the band edge.
To understand the implications of these features, 
the inverse screening length is plotted,
as a function of the momentum transfer, $q$, 
for different values of the Fermi energy
in Fig.~\ref{fig:pi}(b-c) and 
for a fixed carrier density in Fig.~\ref{fig:pi}(d). 
%

Fig. \ref{fig:pi}(b) shows 
the inverse screening length of the Mexican hat dispersion
at 3 different temperatures with 
the Fermi level fixed at $5$ meV above the band edge. 
The low--temperature ($T=5$ K) curve has a strong $q$ dependence 
that arises from the bandstructure.
There are two Fermi wavevectors for $0 < E_F < \epsilon_{0}$
denoted as $k_{F_1}$ and $k_{F_2}$ and 
illustrated in the inset of Fig. \ref{fig:pi}(c).
The two Fermi wavevectors result in three features 
for $\Pi (q) $ at $T=5$ K.
These features correspond to momentum transfers of
$ q = 2k_{F_1} $,
$ q = 2k_{F_2} $, and
$ q = k_{F_2} - k_{F_1}$. 
Just as with the parabolic dispersion, 
there is a sharp change in the derivative of $\Pi(q)$ 
when $q$ is twice the Fermi wavevector,
except now there are two Fermi wavevectors. 
The third and largest peak occurs when $q = k_{F_2} - k_{F_1}$, 
which is the minimum momentum required to 
transfer between the two Fermi rings.
This can be viewed as a type of Fermi surface nesting.
Increasing the temperature smooths out these sharp features, and 
at $T=300$ K, $\Pi(q)$ smoothly decreases with increasing $q$.
When the Fermi level is $3$ meV below the band edge as in Fig. \ref{fig:pi}(c),
the screening at $T=5$ K is essentially zero since there are no carriers,
and the qualitative features of the polarization functions at 
$77$ K and and $300$ K are 
the same as those in Fig. \ref{fig:pi}(b)  
with a small reduction in the overall magnitude resulting from 
the reduced carrier density. 

Fig. \ref{fig:pi}(d) shows the inverse screening lengths at a fixed carrier 
density of $ 10^{13}$ cm$^{-2}$ for different temperatures.
Now, the Fermi level moves with temperature as shown in the legend.
At $5$ K, the Fermi level is $1.6$ meV above the band edge, and 
the small $q$ peak becomes very large as the Fermi level approaches 
the $1/\sqrt{E}$ singularity in the density of states.
At $77$ K and $300$ K, 
the Fermi levels are in the band gap, and 
the polarization functions are similar to those in Fig. \ref{fig:pi}(c).

\begin{figure}
\includegraphics[width=\linewidth,keepaspectratio]{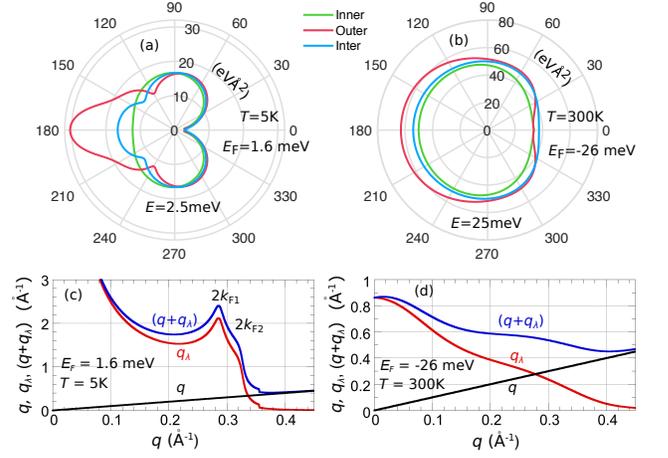}
\caption{
\label{fig:crt_exp}
Polar plots of the matrix elements of
the RPA screened Coulomb potential as a function
of scattering angle at (a) $T=5$ K and $E=2.5$ meV
and (b) $T=300$ K and $E=25$ meV.
The polar angle $\theta$ is the angle between $\kv$ and $\kv'$.
The legend refers to the 3 curves in each polar plot. 
``Inner'' denotes matrix elements with $\kv$ and $\kv'$ both on the inner ring,
``Outer'' denotes matrix elements with $\kv$ and $\kv'$ both on the outer ring,
and
``Inter'' denotes matrix elements with $\kv$ on the inner ring and $\kv'$ on the outer ring.
(c) $q$, $\ql$, and $q+\ql$ as a function of $q$ corresponding to (a). 
(d) $q$, $\ql$, and $q+\ql$ as a function of $q$ corresponding to (b). 
The carrier density is fixed at $10^{13}$ cm$^{-2}$ for all figures.
}
\end{figure}

Now, we consider the magnitude and angle dependence
of the matrix elements $\bra{\kv} \tilde{V} \ket{\kv'}$
of the screened Coulomb potential, 
given by Eq. (\ref{eq:coulomb_pot2})
with $q = |\kv - \kv'|$ and $z = z_0$.
Fig. \ref{fig:crt_exp} shows polar plots of the screened Coulomb potential 
with $n_s = 10^{13}$ cm$^{-2}$ at two different temperatures and energies.
%
%
The polar angle $\theta$ is the angle between $\kv$ and $\kv'$.
The relevant $\ql$ plots are shown in Fig. \ref{fig:pi}(d).
For a fixed energy,
scattering can occur within the inner ring ($\kv$ and $\kv'$ both lie on the inner ring),
within the outer ring ($\kv$ and $\kv'$ both lie on the outer ring),
or between the inner ring and the outer ring ($\kv$ and $\kv'$ lie on different rings).
These 3 different matrix elements are denoted `Inner,' `Outer,' and `Inter,' respectively,
in Fig. \ref{fig:crt_exp}.

We first consider the low-temperature $T=5$ K matrix elements
at an energy of 2.5 meV above the band edge
shown in Fig. \ref{fig:crt_exp}(a).
At the carrier density of $n_s = 10^{13}$ cm$^{-2}$, $E_F = 1.6$ meV, 
The wavevector dependent screening $\ql$
corresponds to the upper curve in Fig. \ref{fig:pi}(d), and a
more detailed view is shown in Fig. \ref{fig:crt_exp}(c).
At $E = 2.5$ meV,
the radius of the inner ring $k_1 = 0.142 \: {\rm \AA}^{-1}$, 
the radius of the outer ring $k_2 = 0.165 \: {\rm \AA}^{-1}$,
and $k_2 - k_1 = 0.023 \: {\rm \AA}^{-1}$.
At $\theta = 0^\circ$, $q=0$ for the inner and outer ring matrix elements
and $q=k_2 - k_1 = 0.023 \: {\rm \AA}^{-1}$ for the inter ring matrix element.
At $q=0$, $\ql = 9.1 \: \rAAi$, and at $q=0.023 \srAAi$, $\ql = 9.9 \srAAi$.
Thus, at $\theta = 0^\circ$, all three scattering mechanisms are strongly suppressed by
the screening.
%
%
The $\theta = 0^\circ$ inter ring scattering is a backscattering process,
since the two rings have opposite velocities.
Thus, the small $q$ inter ring backscattering is strongly suppressed by the screening.
The values of $q$, $\ql$ and $q + \ql$ are plotted in Fig. \ref{fig:crt_exp}(c).
The value of $\ql$ in the range of $0 \leq q \leq 2 k_{F_2}$ is much larger than $q$.
This means that for $q \leq 2k_{F_2}$, 
the $q$ dependence of $V(q)$ is determined solely by the $q$ dependence of the polarization, 
and the bare momentum transfer $q$ is negligible in comparison. 

Since $\ql$ falls rapidly as $q$ increases, the RPA screened Coulomb potential
in a Mexican hat bandstructure favors large angle scattering.
This is opposite to the trend resulting from the bare $1/q$ Coulomb interaction.
The large outer-ring matrix elements for $\theta$ between $150^\circ$ and $210^\circ$
arise because the momentum transfer around the outer ring becomes larger than $2k_{F_2}$.
The kink at $120^\circ$ corresponds to the peak in $\ql$ at $2k_{F_1}$. 
At low temperature, the polarization strongly suppresses the magnitude of the 
matrix elements at the Fermi level.
Only for those energies several $k_BT$ above the Fermi level can the momentum transfer become
large enough that the polarization becomes negligible, and $V(q)$ returns to a $1/q$ dependence. 
This large momentum transfer corresponds to backscattering across the outer ring.

Fig. \ref{fig:crt_exp}(b) shows the $T=300$ K matrix elements at an 
energy of 25 meV above the band edge.
As the temperature increases to 300 K, both the magnitude and the angular dependence
of the matrix elements change considerably compared to those at $T=5$ K.
This is a result of the large change in the polarization function as shown in 
Fig. \ref{fig:pi}(d).
An enlarged view of the $T=300$ K $\ql$ curve is shown in Fig. \ref{fig:crt_exp}(d).
The Fermi level now lies below the band edge at $E_F=-26$ meV.
Compared to the $T=5$ K polarization, 
the magnitude of the polarization decreases by an order of magnitude at the bandedge,
the sharp features disappear, and $\ql$ monotonically decreases as $q$ increases.  
However, the overall decrease of $q + \ql$ over the range of relevant $q$ values is relatively small.
At $E = 25$ meV, $k_1 = 0.11 \srAAi$ and $k_2 = 0.187 \srAAi$.
At $q=0$, $\qpql = 0.861 \srAAi$, and at $q = 2k_2 = 0.374 \srAAi$, $\qpql = 0.462 \srAAi$.
Thus, the maximum increase in the matrix element going from $\theta=0$ to $\theta = 180^\circ$
is a factor of $1.8$, which is shown for the matrix elements of the outer ring
in Fig. \ref{fig:crt_exp}(b).
Over the entire range of relevent momentum transfer $q$, the $T=300$ K  
polarization is much less than the $T=5$ K polarization, so that the matrix elements are 
uniformly larger at $T=300$ K compared to those at $T=5$ K. 
Since the scattering rate is proportional to $|V(q)|^2$, the scattering rates will
be significantly larger at room temperature compared to those at low temperature.

\begin{figure}
\includegraphics[width=\linewidth,keepaspectratio]{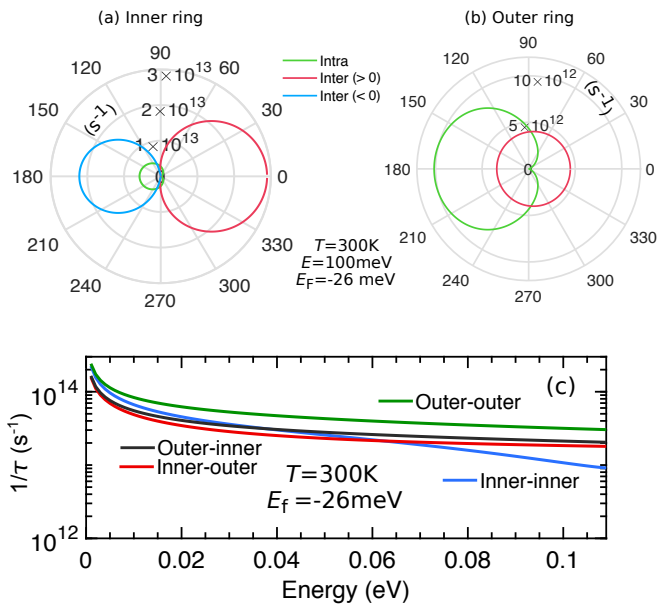}     
\caption{
Polar plots of $1/\tau(\theta)$ given by the integrand of Eq. (\ref{eq:tau})
for (a) $k$ on the inner ring and (b) $k$ on the outer ring.
Inter-ring and intra-ring rates are indicated by the legend. 
Inter-ring contributions can be either positive or negative.
(c) Four components of the total scattering rate as a function of energy.
The components are indicated by the legends where,
for example, ``Inner-outer'' denotes the initial state on the inner ring and the
final state on the outer ring.
\label{fig:S_int} 
}
\end{figure}

The integrand that determines the momentum scattering rates at a given energy $E$, 
given by Eq. (\ref{eq:tau}),
contains not only $|V(q)|^2$, but 
also the final density of states and the relative change in the velocity which can be 
positive or negative.
The $[1 - \frac{v'}{v}\cos(\theta_{\vv, \vv'})]$ term further reduces the small 
angle intra-ring matrix elements, which are already small due to the large polarization at small $q$. 
The integrand of Eq. (\ref{eq:tau}) is plotted in Fig. \ref{fig:S_int} at $T=300$ K, $E_F = -26$ meV,
and $E = 100$ meV.
Fig. \ref{fig:S_int}(a) shows the angle-dependent scattering rate for the initial $k$ on the inner ring,
and Fig. \ref{fig:S_int}(b) shows the angle-dependent scattering rate for the initial $k$ on the outer ring.
Note that the energy $E=100$ meV is $10$ meV below the top of the hat in Fig. \ref{fig:dispersion_dos}(a).
At this energy, the magnitude of the group velocity of a state on the inner ring is much less that
of a state on the outer ring.
For inter ring scattering from the inner ring to the outer ring, $v' > v$, 
and a forward scattering process with $\theta_{\vv,\vv'}=0$
causes the $[1 - \frac{v'}{v}\cos(\theta_{\vv, \vv'})]$ term in the integrand to become negative.
The forward scattering process with $\theta_{v,v'} = 0^\circ$ corresponds to 
backscattering in $k$-space with $\theta = 180^\circ$, where 
$\theta$ is the angle between the initial state $k$ on the inner ring
and the final state $k'$ on the outer ring.
Thus, in Fig. \ref{fig:S_int}(a), 
the negative values of $1/\tau(\theta)$, 
shown by the blue curve, are centered around $\theta = 180^\circ$.
Backscattering with $\theta_{v,v'} = 180^\circ$ corresponds to forward scattering
in $k$-space with $\theta = 0^\circ$, and the corresponding positive values of 
$1/\tau(\theta)$ are shown by the red curve centered around $\theta = 0^\circ$. 
When scattering from the outer ring to the inner ring, $v'/v < 1$, so that 
$1/\tau(\theta)$ is positive for all angles as shown in Fig. \ref{fig:S_int}(b). 

Fig. \ref{fig:S_int}(c) shows the 4 components of the total scattering
rate as a function of energy at $T=300$ K. 
The energy $100$ meV corresponds to the polar plots shown in (a) and (b).
As the energy approaches the top of the hat, $110$ meV,
the radius $k_1$ of the inner ring goes to zero, 
so that $q = |\kv_2 - \kv_1|$
becomes independent of $\theta$.
The denominator in Eq. (\ref{eq:coulomb_pot2}) is then independent of $\theta$, 
the $\cos(\theta_{v,v})$ term integrates to zero,
and the integral over $\theta$ gives $2\pi$.
Thus, in the limit $E$ approaches $\ep_h$ from below, 
the integral in Eq. (\ref{eq:tau}) can be performed analytically
for both inter-ring scattering and intra-ring scattering within the inner ring. 
At $E=\ep_h$, the single-spin density of states of both the inner ring and the outer ring
are equal to $\frac{m^*}{2\pi \hbar^2}$.
For inter-ring scattering, $q = k_2$, and the inter-ring scattering rate is
\begin{align}
\frac{1}{\tau_{\rm inter}}&= \frac{n_I e^4}{4 \ep^2 \hbar} \frac{m^*}{2 \pi \hbar^2}
\frac{ 2 \pi }{\left(k_2  + \ql(k_2) \right)^2} 
\nonumber \\
& =
4 \pi^2 \alpha^2 \frac{m_0 c^2}{\hbar} \frac{m_r}{\ep_r^2} \frac{ n_I }{\left(k_2  + \ql(k_2) \right)^2}
\nonumber \\
& = 2.06 \times 10^{13} \; {\rm s^{-1}},
\label{eq:tau_eh}
\end{align}
where $k_2 = 2 \sqrt{2 m^* \ep_h}/ \hbar = 0.217 \: {\rm \AA}^{-1}$
and $\ql(k_2) = 0.364 \: {\rm \AA}^{-1}$.
In the second line of Eq. (\ref{eq:tau_eh}), $\alpha$ is the fine structure constant,
$m_0$ is the bare electron mass, $c$ is the speed of light, $m_r = 0.409$ is the relative mass,
and $\ep_r = 3.1$ is the relative dielectric constant.
For intra-ring scattering within the inner ring,
$q \rightarrow 0$, and the intra-ring scattering rate becomes
\begin{equation}
\frac{1}{\tau_{\rm intra}} =
4 \pi^2 \alpha^2 \frac{m_0 c^2}{\hbar} \frac{m_r}{\ep_r^2} \frac{ n_I }{ q_{\lambda}^2(0) }
= 9.38 \times 10^{12} \; {\rm s^{-1}},
\label{eq:tau_eh_intra1}
\end{equation}
where $\ql(0) = 0.861 \; {\rm \AA}^{-1}$.
The reduction of $1/\tau_{\rm intra}$ with respect to $1/\tau_{inter}$
is solely the result of the increased value of $\ql$ as $q \rightarrow 0$.
The largest component to the total scattering rate is from scattering
within the outer ring.
Scattering within the outer ring allows for the largest momentum
transfer $q$ and thus the smallest values of $\ql$.
%

%
\begin{figure}
\includegraphics[width=\linewidth,keepaspectratio]{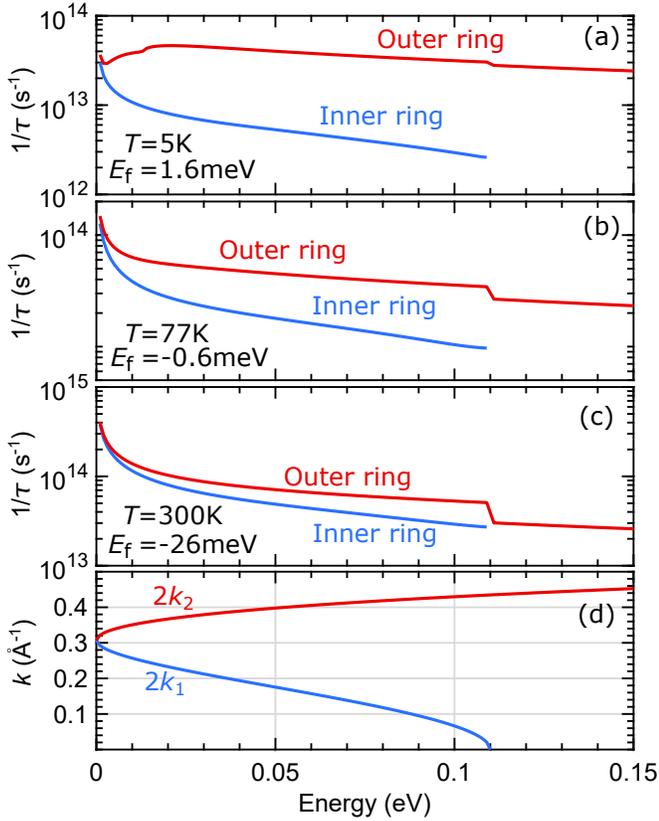}     
\caption{
Momentum scattering rates for a charge density of $ 10^{13}$cm$^{-2}$ at
(a) $5$ K, (b) $77$ K and (c) $300$ K. 
The band parameters correspond to the ones used in 
Fig. \ref{fig:dispersion_dos}, and the polarization functions are shown in Fig. 
\ref{fig:pi}(d).
(d) Maximum values of $q$ for scattering from the outer ring ($2k_2$) 
or within the inner ring ($2k_1$) as a function of energy. 
\label{fig:crt_gen} 
}
\end{figure}

The total scattering rates for an initial state on the inner or the outer ring
are shown in Fig. \ref{fig:crt_gen} for the same 
charge density ($10^{13}\;{\rm cm}^{-2}$) and temperatures as in Fig. \ref{fig:pi}(d).
The parameters are also the same as the ones used in the calculation of the screened
Coulomb matrix elements in Figs. \ref{fig:crt_exp} and \ref{fig:S_int}. 
At $T=5$ K, as a result of the extremely large polarization,
the scattering rate is suppressed for energies below 14 meV. 
At $E=14$ meV, $2k_2 = 0.36 \; {\rm \AA}^{-1}$.
At energies below 14 meV, the polarization is large for all possible 
momentum transfer $q$, the matrix elements of the screened Coulomb potential
are reduced, and the scattering rate is reduced.
The low-energy minimum occurs at $E=2.5$ meV, when the minimum inter-ring scattering
momentum $q = 0.023 \; {\rm \AA}^{-1}$ is where the polarization function has its maximum value.
As the energy decreases below $2.5$ meV towards the band edge, the $1/\sqrt{E}$ density of states term
in Eq. (\ref{eq:tau}) takes over, and the rate increases as $E \rightarrow 0$.
For momentum transfer $q \gtrsim 0.36 \; {\rm \AA}^{-1}$, the polarization is negligible,
and the RPA screened potential reverts to the bare unscreened potential as shown in 
Fig. \ref{fig:crt_exp}(c).
As the energy increases above 14 meV, unscreened backscattering takes place within the 
outer ring. 
The energy dependence for higher energies is governed by the energy dependence of the density of states
and the $1/q^2 \approx 1/4k_2^2$ dependence of the matrix element squared.

The radius $k_1$ of the inner ring is maximum at $E=0$ and decreases with increasing energy.
Thus, the polarization relevant to the inner ring matrix elements increases with energy, 
causing the matrix elements to decrease. 
The density of states monotonically decreases and the scattering rate for states on the inner ring
monotonically decreases with energy.
The total rate is dominated by the intra-ring scattering of the outer ring. 

At $T=77$ K, the polarization loses its sharp features and its magnitude is everywhere reduced
causing an overall increase of the scattering rates
and a monotonic decrease with energy. 
This trend is more pronounced at $T=300$ K where there is relatively little change in the sum
$q + \ql$ over the range of relevant energies, and the energy dependencies of the rates
are determined by the $1/\sqrt{E}$ dependence of the density of states.

\begin{figure} 
	\includegraphics[width=\linewidth,keepaspectratio]{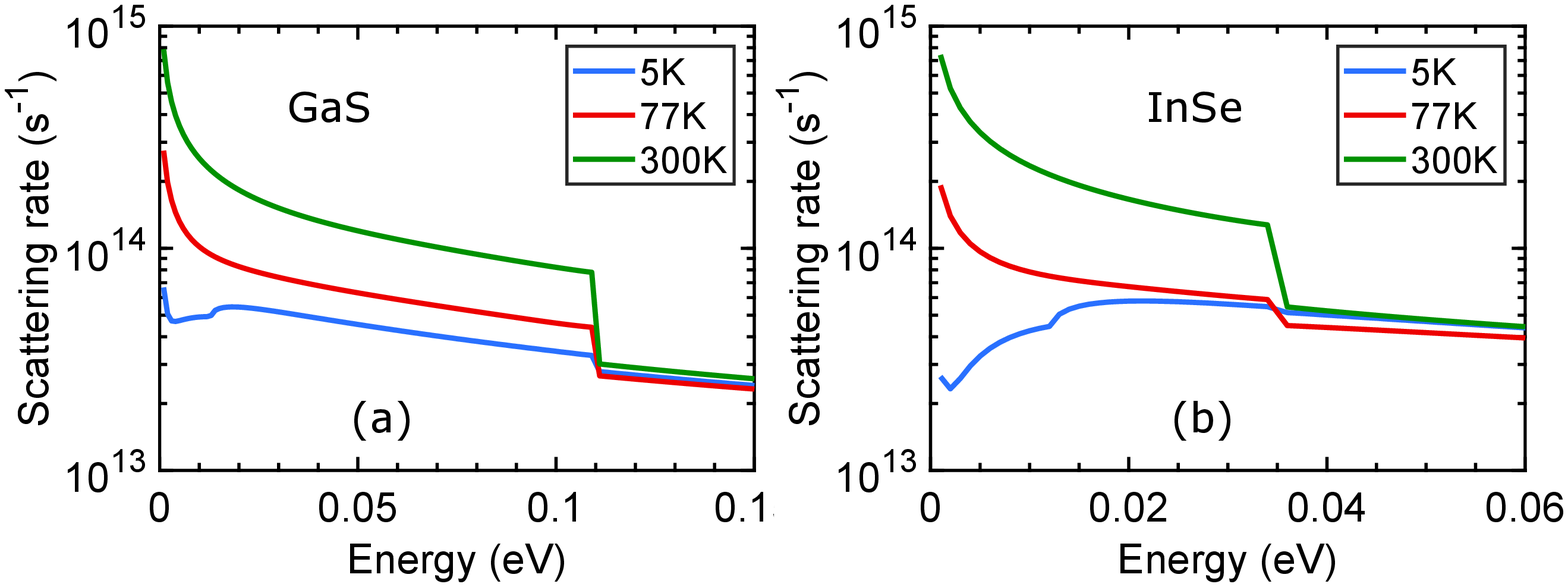}     
\caption{\label{fig:crt_mat} 
Energy dependence of the total momentum relaxation rates 
for (a) GaS and (b) InSe for 3 different temperatures. 
The charge density is fixed at $10^{13}$cm$^{-2}$. 
Parameters used for GaS and InSe are tabulated in	Table \ref{tab:mat_params}.
	}
\end{figure}

The total scattering rates for GaS and InSe are shown in Fig. \ref{fig:crt_mat} 
for temperatures of $5$ K, $77$ K and $300$ K.
The temperature dependence of the overall magnitudes of the scattering rates 
are determined by the magnitudes of the
matrix elements squared of the screened Coulomb potential, which, in turn, are determined by the temperature 
dependence of the polarization functions as shown in Figs. \ref{fig:pi}(d) and \ref{fig:crt_exp}.
When the energy is equal to the height of the hat, 
the contribution from the inner-ring scattering disappears giving an abrupt decrease
in the total scattering rate at $T=77$ K and $300$ K. 
At $T=5$ K, the scattering rate from the inner ring is always small compared to that of the outer ring
(except right at the band edge), so that the small discontinuity at $E=\ep_h$
is primarily the result of the disappearance of the inter-ring scattering from the outer ring
to the inner ring.
For energies above the top of the hat,  
the rates become almost identical differing by at most a factor of 1.2 for InSe.
The fine differences result from the details of the different Fermi levels combined with the different thermal 
broadening for each different temperature.
The large decrease in the $T=5$ K, low-energy
scattering rate for InSe compared to GaS is the result of the larger polarization
in InSe due to its larger mass and larger density of states.

\begin{figure}
\includegraphics[width=\linewidth,keepaspectratio]{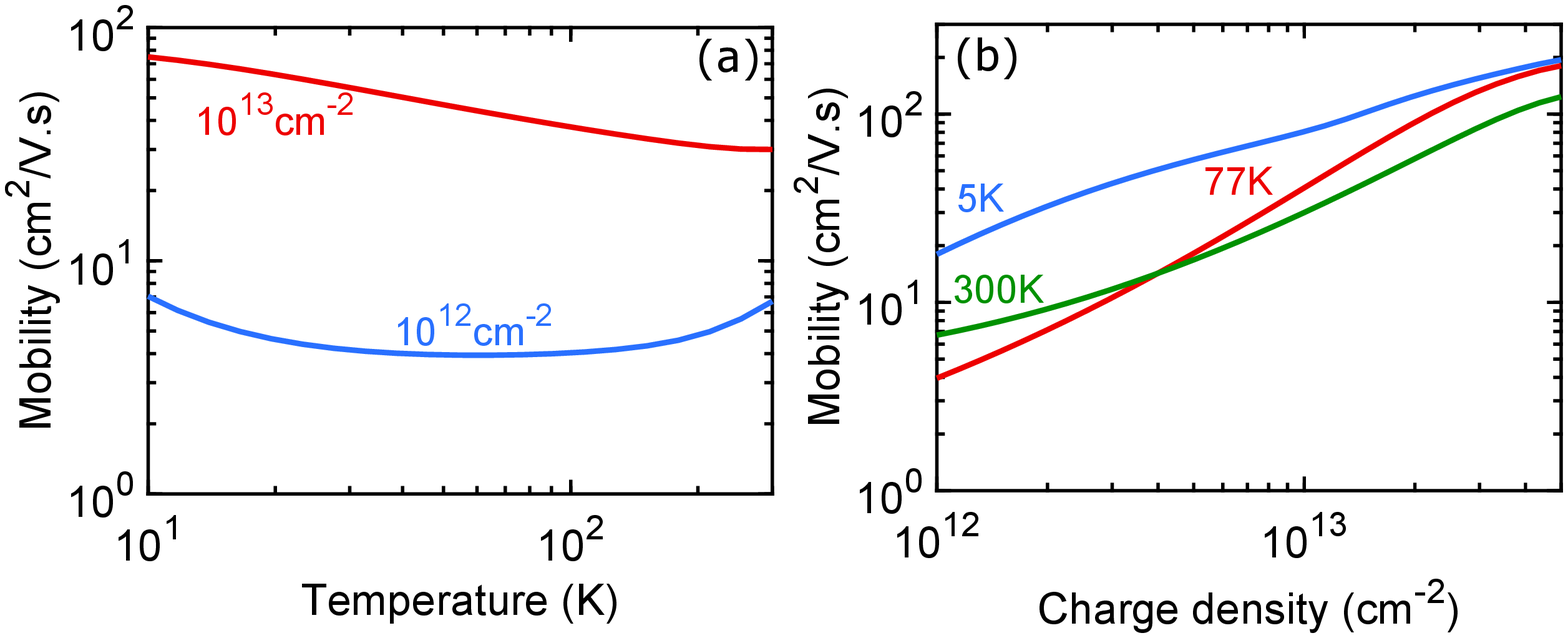}
\caption{
\label{fig:mu_rpa} 
Charged impurity limited hole mobility of GaS  
(a) as a function of temperature for a carrier density of 
10$^{12}$ cm$^{-2}$ (blue) and 10$^{13}$ cm$^{-2}$ (red) and
(b) as a function of carrier density at $5$ K (blue), $77$ K (red) and 
$300$ K (green) for a fixed charge density of 10$^{13}$ cm$^{-2}$.
The impurity density $n_I$ is fixed at $10^{12}$ cm$^{-2}$.
}	
\end{figure}

The temperature and charge density dependence of the mobility
are plotted in Fig. \ref{fig:mu_rpa}.
Both the temperature dependence and the density dependence of the mobility
are primarily governed by the temperature and density dependence of the polarization.
The initial decrease in mobility with temperature results from the decrease in
polarization with temperature as shown in Fig. \ref{fig:pi}(d). 
The decrease in screening, increases the matrix element squared which increases the scattering rate
and decreases the mobility.
At $T=300$ K, there is a significant contribution to the integrand ($\mu(E)$) of Eq. (\ref{eq:mu}) from
energies above $\ep_h$. 
Once $E = \ep_h$ starts to fall inside the thermal window defined by $-\partial f_0/\partial E$
in Eq. (\ref{eq:mu}), the sudden decrease in $1/\tau(E)$ shown in Fig. \ref{fig:crt_mat},
gives rise to a corresponding increase in $\mu(E)$, so that the integral begins
to increase with temperature.
The `turn-on' or `thermal activation' of the mobility starts to be seen at lower temperatures for
lower carrier densities as shown in Fig. \ref{fig:mu_rpa}(a).
For lower carrier densities, screening is less, the matrix elements and scattering rates are larger
at lower energies,
the low-energy values of $\mu(E)$ are reduced, and the discontinuity at $E=\ep_h$ is larger
so that the higher energies give a disproportionally larger contribution to the mobility. 

For a fixed temperature, 
as the charge density increases, the screening increases, which reduces the matrix element squared
and the scattering rates and increases the mobility as seen in Fig. \ref{fig:mu_rpa}(b).
At a charge density of $5 \times 10^{13}$ cm$^{-2}$, the mobility is between $100 - 200$ cm$^2$/V$\cdot$s 
for the 3 temperatures, 5 K, 77 K, and 300 K.

\begin{table}[b]
    \centering
    \begin{tabular}{c   c  c c}
        \hline
        \hline
        \multirow{2}{*}{Material}\Tstrut  & 
        \multirow{2}{*}{\begin{tabular}{@{}c@{}}Effective mass \\ m* (m$_0$) 
        \end{tabular}}  & \multirow{2}{*}{\begin{tabular}{@{}c@{}}Height of the 
        hat\\ ($\epsilon_h$)  (meV) \end{tabular}} &  \multirow{2}{*}{\begin{tabular}{@{}c@{}}Relative\\ permittivity \end{tabular}}       \\
        & & \\ [0.5ex]
        \hline
        GaS\Tstrut   & 0.409  &  111.2 & 3.10  \\ [0.5ex]
        GaSe         & 0.600  &  58.7  & 3.55 \\ [0.5ex]
        InS          & 0.746  &  100.6 & 3.08 \\ [0.5ex]
        InSe         & 0.926  &  34.9  & 3.38 \\ [0.5ex]
        \hline
        \hline
    \end{tabular}
    \caption{Effective mass and height of the hat for III-VI materials with
        Mexican hat \cite{darshanamhat}.}
    \label{tab:mat_params}
\end{table}

\begin{figure}
\includegraphics[width=\linewidth,keepaspectratio]{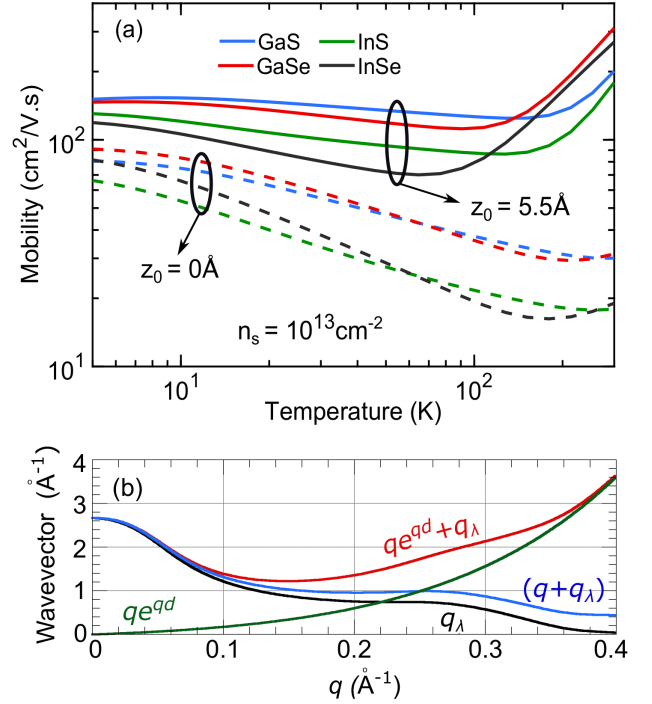}
\caption{
(a) Charged impurity limited monolayer hole mobility as a function of
temperature of 
GaS, GaSe, InS and InSe with the charged impurities 
in the middle of the channel ($d = 0$, dashed line) 
and on the substrate ($d=5.5 \; {\rm \AA}$, solid line).
(b) $qe^{qd}$, $\ql$, $q + \ql$, and $qe^{qd} + \ql$ for GaS at $T=77$ K
where $d = 5.5$ \AA.
The hole density $n_s = 10^{13}$ cm$^{-2}$, and
the charged impurity density $n_I$ is fixed at $10^{12}$ cm$^{-2}$.
\label{fig:mu_vs_T} 
}	
\end{figure}

\begin{figure}
\includegraphics[width=\linewidth,keepaspectratio]{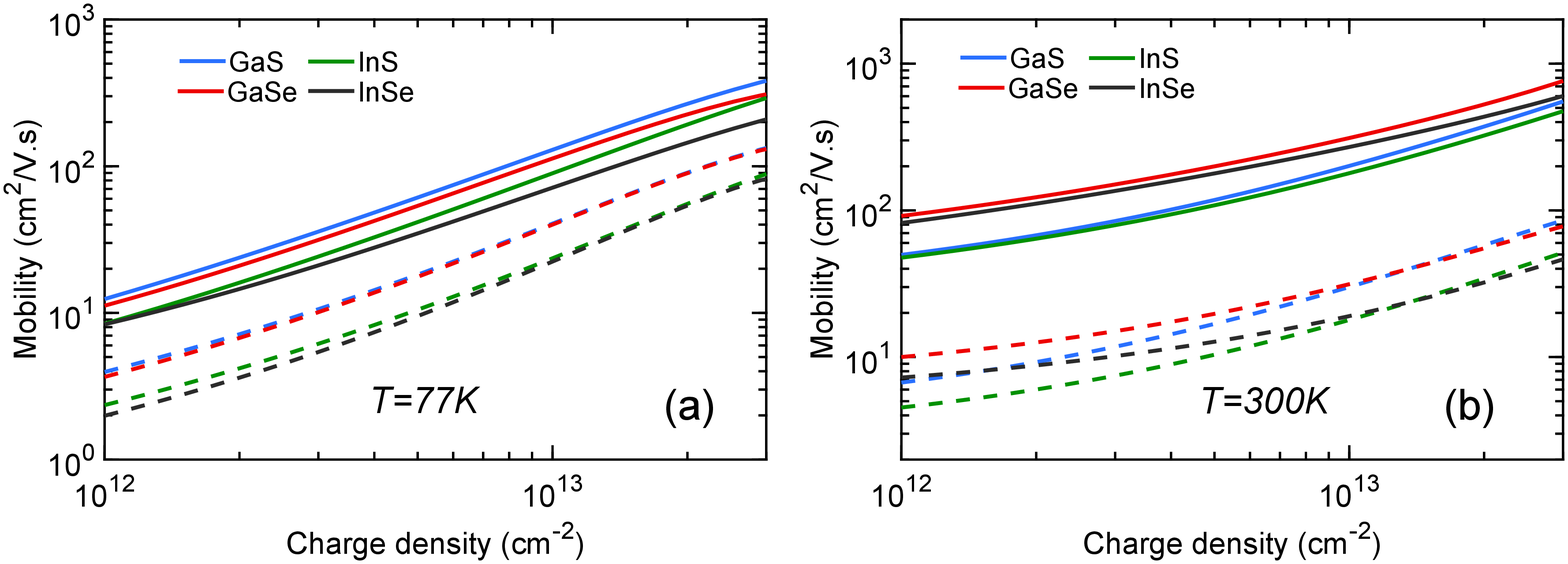}
\caption{
Charged impurity limited monolayer hole mobilities as a function of carrier density at 
(a) $T=77$ K and (b) $T=300$ K for the 4 III-VI materials as indicated by the legends.
Solid lines result from charged impurities on the substrate ($z_0 = 5$ {\AA}),
and the dashed lines result from charged impurities in the middle of the channel
($z_0 = 0$).
The impurity density $n_I$ is fixed at $10^{12}$ cm$^{-2}$.
\label{fig:mobility_vs_density} 
}	
\end{figure}

The temperature dependence of the 4 III-VI p-type materials are shown in Fig. (\ref{fig:mu_vs_T})
for a fixed hole density of $10^{13}$ cm$^{-2}$ and two different positions of the
charged impurities, in the middle of the channel ($z_0 = 0 $\AA) and on the substrate ($z_0 = 5.5$ \AA).
The relevant material parameters are given in Table \ref{tab:mat_params}.
The general trends of the temperature dependence follow those seen in Fig. \ref{fig:mu_rpa}.
The low temperature mobilities order according to the effective masses with the lower masses 
correlating with the higher mobilities.
However, the dependence is weaker than a $1/m^*$ dependence.
The minimum and maximum effective mass differ by a factor of 2.3, and
at $T=5$ K, the mobilities differ by a factor of 1.4.
The difference in mobilities increases to a maximum of 2 near the beginning
of the high-temperature crossover where the mobilities start to increase.
The cross-over begins at a lower temperatures for the materials with a smaller 
value of $\ep_h$, since lower temperatures can thermally excite carriers above the top of the hat.

Moving the charged impurities from the middle of the channel to the substrate
increases the mobility, as would be expected, since the charged impurities are further
away from the carriers.
However, it also lowers the temperature of the high-temperature crossover, which is not an
obvious consequence. 
The reason lies in the large enhancement of the bare, large-wavevector screening as shown in 
Fig. \ref{fig:mu_vs_T}(b) for GaS at $T=77$ K.
For GaS, the bare term $qe^{qd}$ in the denominator becomes larger than $\ql$
at $q=0.22 \; {\rm \AA}^{-1}$.
The minimum value of $2k_2$ is $0.31 \; {\rm \AA}^{-1}$ at the band edge,
and at the top of the hat, $2k_2 = 0.43 \; {\rm \AA}^{-1}$.
At that value of $q$, the denominator $qe^{qd} + \ql$ is larger
than at $q=0$, so that backscattering
across the outer ring is strongly suppressed
giving a large enhancement to $\mu(E)$ for energies $E=\ep_h$.

The hole density dependence of the charged impurity limited 
mobility at $T=77$ K and $T=300$ K is shown in Fig. (\ref{fig:mobility_vs_density}).
The mobility monotonically increases with hole density $p_s$ for
a fixed charged impurity density $n_I$.
This trend would be expected due to increased screening resulting from the higher hole density.
At the highest hole densities considered of $3 \times 10^{13}$ cm$^{-2}$, with the
charged impurities on the substrate, the 
$T = 300$ K mobilities lie between 500 and 800 cm$^2$/V$\cdot$s for all 4 materials.
With the impurities at the center of the channel, the mobilities decrease
one order of magnitude and lie in the range of 50 to 80 cm$^2$/V$\cdot$s. 
All mobilities are calculated for a charged impurity density of $n_I = 10^{12}$ cm$^{-2}$, 
and the mobilities are inversely proportional to $n_I$, so that all mobility values shown
can be easily scaled for arbitrary values of $n_I$. 

\section{Conclusions}
The Mexican hat type bandstructure that occurs in the valence band of monolayer
and few layer III-VI materials and other 2D materials 
gives rise to unique screening properties.
The singular density of states
at the band edge and the two Fermi wavevectors up to the height of the hat,
lead to large screening and strong wavevector dependence of the screening.
The wavevector dependence of the screened Coulomb interaction is so strong, that 
the temperature and density dependence of the matrix element squared
is the dominant factor determining the overall trends with respect
to temperature and density.
The reduction of polarization with temperature causes an initial increase
in scattering and decrease in mobility with increasing temperature.
%
Short wavevector inter-ring backscattering and scattering
within the smaller ring is always suppressed by the
large polarization at small $q$.
When the the charged impurities lie in the middle of the 2D channel,
the wavevector dependence of the polarization 
favors large wavevector backscattering across the outer ring.
When the charged impurities lie on the substrate, the bare
screening increases rapidly at larger wavevectors suppressing
the backscattering within the outer ring.
For charged impurities on the substrate, the polarization suppresses
the small wavevector scattering and the exponential wavevector dependence of the bare 
Coulomb interaction suppresses the large wavevector scattering across the outer ring
leading to an overall increase in mobility.
The suppression of the large wavevector scattering also 
reduces the temperature at which the mobility starts to increase when the charged 
impurities are on the substrate.
The mobility monotonically increases with hole density up to the maximum
value considered of $3 \times 10^{13}$ cm$^{-2}$ where it reaches a maximum
value of 800 cm$^2$/V$\cdot$s for GaSe with the charged impurities located on the substrate.
Placing the impurities in the center of the channel reduces the maximum value by
an order of magnitude.
All mobility values are calculated for a charged impurity density of $n_I = 10^{12}$
cm$^{-2}$ and scale inversely proportionally to $n_I$.

\noindent
\begin{acknowledgements}
We acknowledge helpful discussion with Dr. Yafis Barlas.
This work was supported by FAME, one of six centers of STARnet, 
a Semiconductor Research Corporation program sponsored by MARCO and DARPA.
This work used the Extreme Science and Engineering Discovery
Environment (XSEDE), which is supported by National
Science Foundation Grant No. ACI-1548562 and allocation
ID TG-DMR130081. 
\end{acknowledgements}


\end{document}